\documentclass[aps,nofootinbib,longbibliography,prd,twocolumn]{revtex4-1}
\usepackage[babel]{csquotes}
\usepackage{graphicx}
\usepackage{amsmath,amssymb}
\usepackage[colorlinks,citecolor=blue,linkcolor=blue,urlcolor=blue]{hyperref}
\usepackage{mathrsfs}
\usepackage{enumerate}
\def\be{\begin{equation}}
\def\ee{\end{equation}}

\newcommand{\R}{{\scriptstyle  R}}

\begin{document}

\title{On the possibility of laboratory evidence for quantum superposition of geometries}

\author{Marios Christodoulou${}^a$, Carlo Rovelli${}^b$ \vspace{1mm}}

\affiliation{\small
 \mbox{Department of Physics, Southern University of Science and Technology, Shenzhen 518055, China }\\ 
\mbox{CPT, Aix-Marseille Universit\'e, Universit\'e de Toulon, CNRS, F-13288 Marseille, France.}
}
\date{\small\today}

\begin{abstract}

\noindent 
We analyze the recent proposal of measuring a quantum gravity phenomenon in the lab by entangling two nanoparticles gravitationally.  We give a generally covariant description of this phenomenon, where the relevant effect turns out to be a quantum superposition of proper times.  We point out that if General Relativity is assumed to hold for masses at this scale, measurement of this effect would count as evidence for {\em quantum superposition of spacetime geometries}. This interpretation addresses objections appeared in the literature.  We observe that the effect sheds light on the Planck mass, and argue that it is very plausibly a real effect. 
\end{abstract}

\maketitle

\section{Introduction}

An experiment aimed at measuring a quantum gravitational effect in the lab has been recently proposed by Bose \emph{et.al.\,}\cite{Bose2017a} and by Marletto and  Vedral \cite{Marletto2017}. Measurement of the Bose-Marletto-Vedral (BMV) effect may turn out to be a game changer in the tentative field of quantum gravity phenomenology (see the contributions in \cite{Hossenfelder2017} and references therein.)

Consider two particles of mass $m$ brought at a (small) distance $d$ for a time $t$.  The phase $e^{-i\frac{Et}{\hbar}}$ of the quantum state of the particles rotates by the angle $\phi=Et/\hbar$ and the gravitational effect of each particle on the energy $E$ of the other is $\delta\!E=Gm^2/d$. Therefore the time passed by the particles near one another produces a phase shift 
\be
       \delta \phi = \frac{\delta\!E\, t}{\hbar} = \frac{Gm^2t}{\hbar d}
\label{uno}
\ee
in their quantum state.  Equivalently 
\be
       \delta \phi = \alpha\ \Big(\frac{m}{m_{Planck}}\Big)^2, 
\label{due}
\ee
where $\alpha=\frac{ct}{d}$ is a dimensionless parameter characterising the setting and $m_{Planck}=\sqrt{\hbar c/G}$ is the Planck mass, which is of the order of micrograms. 

\begin{figure}[b]
\includegraphics[height=3.1cm]{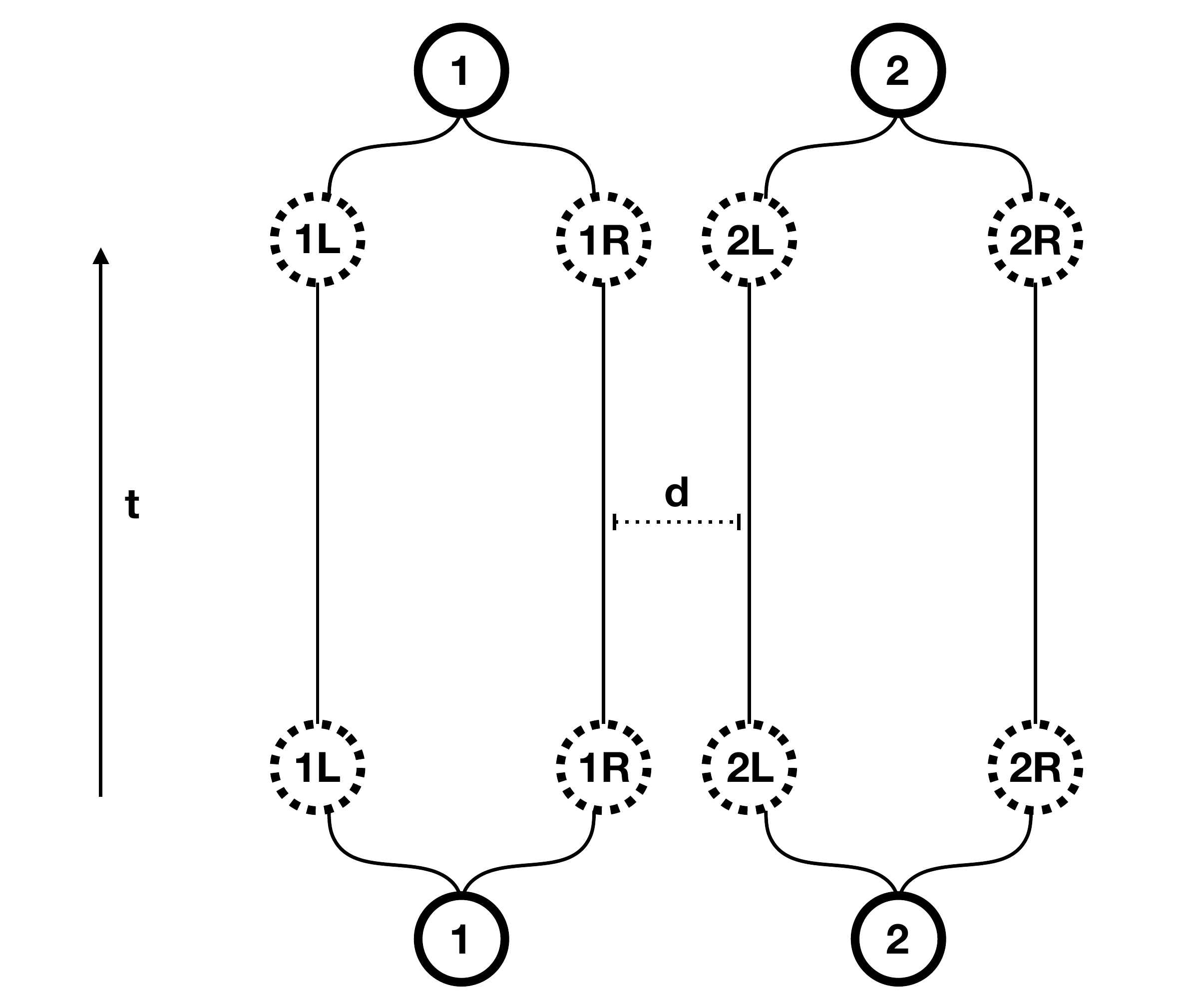}
\caption{The BMV setting.}
\label{uno}
\end{figure}

It is technically possible to split the quantum state of a nanoparticle ($m\sim 10^{-11}gr$) with (embedded) spin into a superposition of two components where the particle is located at different positions ---as in a Stern-Gerlach setting--- and then recombine the two ``branches". It was pointed out in \cite{Bose2017a} that it may soon be possible to split in this manner \emph{two} particles, and keep them nearby ($d\sim 10^{-4} cm$) in only one of the four (two per particle) branches, for a time ($\sim 1 s$) sufficient to reach $\delta\phi\sim \pi$. See Figure 1. In such a configuration, the phase is shifted significantly in this branch alone,  entangling the quantum states of the two particles. Entanglement can then be revealed by checking Bell-like correlations in subsequent spin measurements.  
This (and similar \cite{Marletto2017,Carney2018}) gravitationally mediated entanglement is the BMV effect. 

As emphasized in \cite{Marletto2017,Marletto2017a}, a general argument based on information theory demands that a physical entity can entangle two systems only if it is itself described by quantum (non-commuting) variables, therefore detection of the BMV effect counts as evidence that gravity is quantised (see also \cite{Giampaolo2018,Altamirano2018}.) 

We argue here that, specifically, assuming General Relativity to hold for masses at this scale, detecting the effect counts as evidence that the gravitational field can be in a superposition of two macroscopically distinct classical fields. Since the gravitational field is the geometry of spacetime (measured by rods and clocks), the BMV effect counts as evidence that \emph{quantum superposition of different spacetime geometries is possible, can be achieved in the lab, and has observable effects}.

We present some general considerations on this experiment, give a general relativistic account of it, and comment on some objections to its interpretation that have appeared in the literature.  A general relativistic treatment is not \emph{required} for the effect, because first order perturbative quantum gravity is sufficient, but it clarifies the physical significance of the effect. In particular it shows that the effect reveals the quantum  superposition of two distinct proper times along the worldline of a particle.   

\section{General considerations}

\subsection{A plausible nonrelativistic quantum gravitational effect}

The speed of light $c$ does not enter Eq.~\eqref{uno}.  Hence the BMV effect survives in the non-relativistic limit $c\to\infty$.  Detecting it would not test the fully quantum general relativistic regime, but only the regime where $G$ and $\hbar$ are kept finite while $c$ can be approximated by $c\to\infty$. 

This is a rarely considered regime, because it does not involve the full complexity of the  \emph{relativistic} quantum dynamics of gravity; but it is interesting because in this regime gravity can still keep its quantum properties.  In particular, physical spacetime geometry can be in quantum superposition of macroscopically distinct classical configurations.  As emphasized by Bose \emph{et.al.\,} and by Marletto and  Vedral, the main reason for the interest of the experiment is precisely to provide direct evidence that gravity is quantized. As shown here, on the assumption that General Relativity continues to hold for masses at the nanoparticle scale, detection of the BMV effect would provide evidence in favour of gravity being quantized in the sense that spacetime geometry obeys the superposition principle. 

The possibility of quantum superposition of geometries is largely given for granted in the quantum gravity research community, but its consequences have never been empirically observed, and is still questioned by isolated voices in the literature (see \cite{Carlip2008, Albers2008, Boughn2009, DYSON2013, Bassi2013, Anastopoulos2015, Bassi2017} and references therein.)

The BMV effect is predicted by first order perturbative quantum gravity. Hence it is predicted by any full quantum gravity theory, such as loop quantum gravity and string theory, expected to match perturbative quantum gravity at low energy.   It is therefore very plausible that the effect is real.   

This fact sharply distinguishes the BMV experiment from numerous other attempts to measure quantum gravity effects in the laboratory, because these generally aim at measuring effects that are far more speculative: not predicted by the main current quantum gravity theories, and plausibly unreal.   The current lack of consensus on the best quantum theory of gravity, indeed, does not mean that all wild options are equally plausible and that we have no reason to have reasonable expectations on the behaviour of Nature in unexplored regimes. It happens sometime that Nature surprises us, hence it may be interesting to check---but far more often than not, Nature is remarkably consistent and predictable.  A reasonable bet is therefore that the BVM effect will turn out to be real, while other searched quantum gravity effects in the lab will not. 

For the same reason, on the other hand, measuring the BMV effect is likely not going to be informative about the full high-energy behaviour of quantum gravity and is not going to discriminate between the main current quantum gravity theories, which are compatible with perturbative quantum gravity at low energy.  However, see \cite{noi}, where it is pointed out that a refined version of the BMV experiment may test a relativistic effect as well.

\subsection{Planck mass}

An intriguing aspect of the BMV proposal is that it sheds light on the theoretical meaning of the Planck mass $m_{Planck}$.   While the Planck length and the Planck energy might have a clear physical meaning (the first, as the limit for the physical divisibility of space, the second as the energy where dynamics cannot be anymore described as happening over a spacetime continuum), the physical meaning of Planck mass has remained more elusive.  

Puzzling is the fact that --unlike Planck length and Planck energy-- $m_{Planck}$ falls within a very reachable physical domain: micrograms. It has long been hard to see what sort of quantum gravity effect can happen at the scale of the weight of a human hair.   

Some researchers have suggested that the Planck mass could signal the scale at which quantum theory may break down. After all, the mass of most systems  we treat quantum mechanically is smaller than $m_{Planck}$ and that of most systems we treat classically is larger than $m_{Planck}$. Roger Penrose has suggested that the Planck mass is related to the scale at which the linearity of quantum theory is broken by a physical collapse induced by gravity \cite{Penrose1996}. While logically possible, this is not a straightforward consequence of quantum mechanics and general relativity alone, and it can be viewed as an intriguing but speculative suggestion, not necessarily a clearly plausible consequence of what we know about nature. Interestingly, if Penrose's  suggestion is correct, the BMV effect should presumably \emph{not} happen \cite{Howl:2018qdl,Marletto:2018lsb}, because quantum superposition of macroscopically different spacetimes should be suppressed, and the Penrose collapse time should be of the same order as the BMV time.

But this fact sheds light precisely on what the Planck-mass scale indicates: it is the scale at which quantum superposition of spacetimes curved by masses at this scale may be detectable. This is a way of reading equation \eqref{due}.  The need to control quantum coherence limits the mass of the particles in the experiment to values much smaller than $m_{Planck}$, but the ``long" length of the time $t$, compared to the light travel time $d/c$, compensates for the smallness of the ratio $m/m_{Planck}$, thanks to the fact that the phase shift cumulates in time. That is, $m/m_{Planck}$ is the `small' quantity that determines the physical effect, and $\alpha$ is the `large' multiplicative factor making it measurable. 

A typical interference effect happens at a given scale, which is determined by $\delta\phi\sim\pi$.  If $\delta\phi$ is too small, the interference is negligible and not observable.  But if $\delta\phi$ is too large, given, say, the resolution of the measuring apparatus, then phase average prevails and interference is not visible either.   (To see interference in a two-slit experiment the wavelength must be comparable to the slit size: neither too much larger nor too much smaller.)   Here the Planck mass determines the scale at which interference between superimposed geometries affected by a quantum mass $m$ may be observable.

 Of course, masses smaller than the Planck mass may be used, as in the current proposal for realising the BMV effect which involves masses about a millionth the Planck mass. Smaller masses are easier to set in a path superposition. However, the more removed is the mass used from the Planck mass, the harder the realization of the experiments with regards the other two experimental parameters that determine $\alpha$: per \eqref{due}, to tune $\alpha$ such that $\delta \phi \sim 1$ we must either hold the superposition for longer times  (increase $t$), or, keep the two branches closer (decrease $d$) while still keeping the electromagnetic interaction negligible with respect to gravity. The values proposed in \cite{Bose2017a} for the experimental parameters ($m\sim 10^{-11}gr$, $d\sim 10^{-4} cm$, $t \sim 1s$), precisely strike a compromise that makes detecting the effect just feasible with current technology. 

\section{General covariant treatment of the BMV effect}

\subsection{Classical theory}

The BMV effect can be predicted by describing gravity in the approximation provided by the Newtonian instantaneous force.  The real physical gravitational interaction between the two particles is of course not instantaneous, but this approximation is sufficient because it is valid in the static limit.  Here the static limit is sufficient because the time $t$ during which the cumulative effect on $\delta\phi$ builds-up is much longer than the light travel-time $d/c$ between the two nearby masses which is the time during which the system is not static. Also, the displacement of the particles due to their gravitational attraction itself is entirely negligible.    An accurate analysis of the dynamical aspects of the experiment, and how taking these into account resolves certain apparent conflicts with causality has been recently given in \cite{Belenchia2018}.  To a good approximation, therefore, we can focus on the static phase alone.   Also, since the gravitational field is small, the effect can be reliably computed using perturbation theory around Minkowski background \cite{Bose2017a,Marletto:2018lsb,Carney2018}.  

But to shed full light on the conceptual implications of the BMV effect is far more enlightening to describe it in the full language of general relativity. This  represents our current best understanding of the physical nature of all gravitational phenomena.  It is in these terms that we can see clearly how this effect involves the quantum superposition of different spacetime geometries. This description addresses also some concerns raised by the gauge dependency of the linearised formalism \cite{Anastopoulos2018}. 

Consider a static configuration of two spherical bodies of mass $m$, remaining at distance $d$ in a gravitational field $g$.  Assume the radius $\scriptstyle  R$ of each body to be much smaller than the distance, $\R\!\ll\!d$, but much larger than the Schwarzschild radius: $r_m\!\ll\!\R$ where $r_m=2Gm/c^2$.   Their gravitational field (that is, the corresponding static solution of the Einstein equations) can be approximated by the weak field form of the metric. That is, there is a coordinate system where the line element takes the form 
\be
ds^2=(1+2\phi(\vec x)/c^2) dt^2-d\vec x^2,
\ee
where the Newtonian potential $\phi(\vec x)$ is the sum of the the Newtonian potentials of the two particles:  $\phi(\vec x)=\phi_1(\vec x)+\phi_2(\vec x)$.   For each particle, this is a function of the distance $r$ from the center of the particle
\be
\phi_i(r)=-\frac{Gm}{r}, \ \ \ \ \ \ \ r>\R
\ee
outside the particle itself; and we take it for simplicity to be constant 
\be
\phi_i(r)=-\frac{Gm}{\R} , \ \ \ \ \ \ \ r<\R
\ee
inside the particle ($i=1,2$).  This implies that inside each particle the metric is (approximately, as $\R\ll d$)
\be
ds^2=\left(1-\frac{2Gm}{\R c^2}-\frac{2Gm}{dc^2}\right) dt^2-d\vec x^2.   \label{gf}
\ee
The proper time measured in this geometry by a clock inside each particle during a coordinate time lapse $t$ is
\begin{eqnarray}
s&=&\int_0^t ds = \int_0^t  \sqrt{1-\frac{2Gm}{\R c^2}-\frac{2Gm}{dc^2}} \ dt   \nonumber \\
&&  \ \ \ \ \ \sim  t \left(1-\frac{Gm}{\R c^2}-\frac{Gm}{dc^2}\right). 
\end{eqnarray}
Since $d\gg Gm/c^2=r_m$ the last term is small.  But, as we shall see below, it may still be revealed by interference if $t$ is large enough.  Its contribution to the proper time is 
\be
\delta s = -\frac{Gm t}{dc^2}.
\label{delay}
\ee

\subsection{Superimposing spacetimes}

Let us now analyse what happens in the BMV setting.  There are three physical entities involved: the two particles and the spacetime metric, namely the gravitational field. We assume that at some initial time these are in some initial tensor quantum state, say
\begin{eqnarray}
|\Psi_0\rangle&=& |\psi_1\rangle \otimes  |\psi_2\rangle \otimes  |g\rangle.
\end{eqnarray}
Here $ |\psi_i\rangle$ with   $i=1,2$ are the initial states of the two particles and  $|g\rangle$ is the quantum state of the gravitational field.  We do not need many hypotheses about the states  of the gravitational field, besides ---crucially--- the superposition principle.   We assume that $|g\rangle$ belongs to a Hilbert space that contains semiclassical states that approximate classical geometries $g$, \emph{but also linear superposition of these}.   This is the key property needed to derive the BMV effect. 

The first step of the experiment consists in splitting the state of each particle into the superposition of two semiclassical quantum states 
\be
 |\psi_i\rangle=\frac{|\psi_i^L\rangle+|\psi_i^R\rangle}{\sqrt{2}}
 \label{spin}
\ee
where the particle have different intrinsic properties as well as two different locations.   For concreteness we can think for instance of a spin-$\frac12$ particle with $|\psi_i\rangle$ being an eigenstate of the $z$ component of the spin and $|\psi_i^L\rangle$ and $|\psi_i^R\rangle$ being orthogonal eigenstates of the $x$ component of the spin located in different spacial positions, as in the standard Stern-Gerlach setting. 

For simplicity, we take the unrealistic simplification that the separation can be done very fast, say much faster than the time $d/c$.  Immediately after the spilt, the metric does not yet have time to change significantly and the state becomes
\begin{eqnarray}
|\Psi_1\rangle&=&\frac12\,\Big(|\psi_1^L\rangle+|\psi_1^R\rangle)  \otimes   (|\psi_2^L\rangle+|\psi_2^R\rangle\Big) \otimes  |g\rangle.
 \\
&=&\frac12\, \Big(|LL\rangle+|RR\rangle + |LR\rangle+|RL\rangle\Big) \otimes  |g\rangle. 
\end{eqnarray}
where we have have used the simpler notation $|LL\rangle\equiv |\psi_1^L\rangle  \otimes |\psi_2^L\rangle $.

In a time of order $d/c$ the displacement of the particle produces a disturbance in the gravitational field that propagates at the speed of light to the distance of order $d$ (and past it) modifying $g$ accordingly.  What matters for the resulting metric, which then is again static in the region, is the distance $d$ between the two particles.  This distance is different in each branch.   The gravitational field in \eqref{gf} depends explicitly on $d$, hence the gravitational field itself must become different in the different branches.  

It is important to stress that two metrics defined by \eqref{gf} with two different values of $d$ are \emph{not} diffeomorphic to one another.  Therefore the difference between the two is definitely \emph{not} a gauge difference.  This is important in relation to objections appeared in the literature claiming that only gauge aspects of gravity are involved in this experiment. 

The metrics in different branches represent distinct spacetime geometries. We denote $g_d$ the metric determined by the two particles being at a distance $d$ and call $d_{LL}, d_{LR}, ... $ the distances in the different branches.  Since $d$ differs in each branch, the outcome of this process is different in each branch, giving  
\begin{eqnarray}
|\Psi_2\rangle&=&\frac12\, \Big( |LL\rangle\otimes |g_{d_{LL}}\rangle+|RR\rangle \otimes |g_{d_{RR}}\rangle
\nonumber \\ && \ \ \  
+ |LR\rangle \otimes |g_{d_{LR}}\rangle+|RL\rangle \otimes  |g_{d_{RL}}\rangle\Big),
\end{eqnarray}
or, in compact notation,
\begin{eqnarray}
|\Psi_2\rangle&=&\frac12\, \Big(|LL\,g_{d_{LL}}\rangle+|RR\,g_{d_{RR}}\rangle 
\\ && \nonumber 
+|LR\,g_{d_{LR}}\rangle+ |RL\,g_{d_{RL}}\rangle\Big).
\label{stato1}
\end{eqnarray}
In this state the metric is not semiclassical anymore.  It is in a superposition of macroscopically distinct semiclassical states, entangled with both  particles. 

Say that in a BMV setting the distance $d$ is taken too large compared to $Gm/c^2$ for any significant effect in \emph{three} of the four branches; while in one of the four branches (say $RL$) the two particles are kept at \emph{small} distance $d$.   The proper time along the particles' worldline in this branch is therefore different from the others by the amount \eqref{delay} computed above.  That is, \eqref{delay} is the delay of a clock located on the particles in this branch with respect to the other branches. Now, the time evolution of the quantum state of a particle of mass $m$ is $e^{-i\frac{mc^2 s}{\hbar}}$, where $s$ is proper time, namely the phase is
\be
\phi=-\frac{mc^2 s}{\hbar}
\ee
and the phase difference between this branch and the others is therefore 
\be \label{eq:combineGRQM}
\delta \phi=-\frac{mc^2 \delta s}{\hbar}=\frac{Gm^2 t}{\hbar d}.
\ee
which is precisely the BMV formula equation \eqref{uno}.  This shows that the BMV effect is a direct consequence of gravitational redshift. 

After a time $t$, the state then becomes, up to an irrelevant overall phase
\begin{eqnarray}
|\Psi_3\rangle&=&\frac12\, \Big(|LL\,g_{d_{LL}}\rangle+|RR\,g_{d_{RR}}\rangle \\
&& \nonumber +|LR\,g_{d_{LR}}\rangle+e^{i\frac{Gm^2 t}{\hbar d}} |RL\,g_{d_{RL}}\rangle\Big). 
\label{stato2}
\end{eqnarray}

Next, the two components of each particle are brought back together, and therefore the metric evolves back to the same state in each branch, and the state becomes 
\begin{eqnarray}
|\Psi_4\rangle&=&\frac12\, \Big(|LL\rangle+|RR\rangle  \\
&& \nonumber \hspace{.6cm} + |LR\rangle+ e^{i\frac{Gm^2 t}{\hbar d}}  |RL\rangle\Big)\otimes |g\rangle,  
\end{eqnarray} 
where the particle states are still different because of the internal degrees of freedom.  When the phase reaches the value $\pi$, namely after a time
\be
      t=\frac{\pi \hbar d}{Gm^2},  \label{tempo}
\ee
the state is 
\begin{eqnarray}
|\Psi_4\rangle&=&\frac12\, \Big(|LL\rangle+|RR\rangle + |LR\rangle- |RL\rangle\Big)\otimes |g\rangle.    
\end{eqnarray}
Tracing over the gravitational degrees of freedom and, say, the degrees of freedom of the first particle, gives the density matrix for the second particle
\begin{eqnarray}
\rho &=&| \psi_2^L\rangle\langle \psi_2^L|+ | \psi_2^R\rangle\langle \psi_2^R| , 
\end{eqnarray}
which is obviously not pure.  That is: the states of the particles are entangled.  In the spin case mentioned above in \eqref{spin}, $\rho$ is proportional to the identity operator in the internal space and hence the two particles are \emph{maximally} entangled.

From this perspective the effect is a genuine interference measurement. The quantity measured is $\delta s$ which is very small with respect to $t$ since \eqref{delay} gives
\be
\frac{\delta s}{t}= \frac{r_m}{2d}\ll 1
\ee
where, we recall, $r_m$ is the Schwarzschild radius of the particles. But the  oscillator giving rise to the interference is the phase factor $e^{i\frac{mc^2}{\hbar}s}$ of the quantum state, whose frequency is very high; its period is $\tau=2\pi \frac{\hbar}{mc^2}$.   For interference, we need half period discrepancy namely $\delta s=\tau/2 $, which gives 
\be
 \frac{r_m}{2d} = \frac{\tau}{2t} 
\ee
This is an equality between two very small (relativistic) quantities, but plugging in it the explicit expressions $r_m=\frac{2Gm}{c^2}$ and  $\tau=2\pi \frac{\hbar}{mc^2}$, we see that the two  $c^2$ in the denominators cancel, giving the non-relativistic relation \eqref{tempo}.

Notice that the  location of the particles relative to the laboratory is irrelevant for the effect: what matters is the location of the particles \emph{relative} to one another and their common gravitational field, which is clearly a diffeomorphism invariant notion (it is the physical distance between the two particles).

In \eqref{eq:combineGRQM} we have combined concepts from General Relativity and Quantum Mechanics. In absence of an established theory for quantum gravity, this step requires some explanation. As we saw in the previous subsection, it is a good approximation to work in the weak-field and static limit. This also justifies treating the nanoparticle as a single particle, as we did above. Then, we took the time evolution of the quantum state of the nanoparticle to be affected by gravity, and in particular by time dilation, by using the proper time as the time parameter. This minimal interplay of General Relativity and Quantum Mechanics has been confirmed empirically by the COW experiment \cite{Colella:1975dq} four decades ago, using neutron beams in the gravitational field of the earth. 

There are three assumptions that go into deriving \eqref{eq:combineGRQM}. The first is taken at the classical level: we have assumed that General Relativity continues to hold for gravitational fields sourced by masses at the scale of a nanoparticle ($\sim10^{-11} gr$). The validity of GR at these mass scales regimes has not been verified experimentally and thus needs to be assumed. Then, \eqref{eq:combineGRQM} follows from the two assumptions on the gravity quantum state space stated in the beginning of this subsection: that it contains 1) semiclassical states and 2) superpositions of these. 

We are thus lead to the conclusion stressed in this work: the BMV effect tests, and if detected would count as strong evidence in favour of, the existence of spacetime superpositions. On one hand, the other two assumptions needed to relativistically derive the effect, that GR holds for small masses at the scale of nanoparticles and that the gravity state includes smooth spacetimes such as the one we experience everyday, are mild assumptions. On the other hand, the quantum superposition of geometries is essential for deriving the effect. Each particle's component (say the $R$ component of the particle 1) must be able to fly through \emph{two distinct proper times}, in two different branches ($RL$ and $RR$).  This is precisely what some hypotheses denying the possibility of quantum superposition of macroscopic  geometries consider impossible.  
\section{Objection and replies}

Some papers have questioned the relevance of the BMV effect for quantum gravity and its precise interpretation. 

In \cite{Anastopoulos2018}, the authors argue that ``at the weak-gravity, non-relativistic limit in which these proposed experiments function, the gravitational interaction is determined by the scalar constraint of General Relativity, and not by a dynamical equation for physical degrees of freedom. The relevant gravitational degrees of freedom in the proposed experiments are pure gauge, with no physical content, either classical or quantum. For this reason, they cannot ascertain the quantum nature of gravity."   

The problem with this line of argument is that the weak-field non-relativistic gravity is only an approximation to the true theory.  It is a viable approximation of course, but it does not imply that in the real physical world the gravitational field fails to be a dynamical entity.  The fact that a correlation between two variables is expressed by a constraint does not imply the absence of a physical entity connecting the two variables: the positions of two objects kept at a fixed distance by a stick are related by a constraint, but the constraint \emph{reveals} the reality of the stick, doesn't \emph{contradict} it.  For a stick to be able to establish a correlation that entangles the two objects, the stick itself must be capable of being entangled.   

Similarly, in the approximation where the transfer-time of the information is neglected, the gravitational field (like the stick) correlates the two particles, and the fact that this correlation is expressed by a constraint in a certain approximation does not change the physical fact that the gravitational field must be a quantum entity in order to correlate the particles.  The correlations captured by Newton's law are the manifestation of an underlying entity: the gravitational field, whether or not we treat it in some approximation.  

More importantly, the arguments in \cite{Anastopoulos2018} do not challenge the conclusion that we are stressing in this paper, namely that detection of the BMV effect reveals that spacetime geometry can be in an entangled state with the particles and hence in quantum superposition of distinct classical configurations.  This is because spacetime geometry is defined by clocks, and clocks run at different rates in the different branches.  This direct interpretation of the BMV effect, stressed in this paper, avoids the torn issue of disentangling what is dynamical or what is gauge in gravity. Spacetime geometry is not just determined by the radiative degrees of freedom of gravity: it is also determined by the presence of matter.   We have shown explicitly above that the difference between the metrics in the different branches is \emph{not} pure gauge.   Hence geometry must still be in a quantum superposition of non-gauge equivalent geometries, for the BMV effect to happen. 

In \cite{Hall2018}, the author points out that strictly speaking the BMV effect does not imply that gravity is described by a \emph{quantum} theory, but, in the words of the article, the BMV effect is ``a test or witness of \emph{nonclassical} gravity."  The reason of the subtlety is the theoretical possibility of third options between spacetime being described by standard classical general relativity or a quantum theory having general relativity in its classical limit.  As these third options are more exotic than quantum gravity, the point does not diminish the interest of the BMV experiment, which is not to rule out all possible explanations of nature (these are always infinite): it is to measure an effect cleanly predicted by conventional quantum gravity and not predicted by conventional classical gravity. 

In extreme synthesis: in the non-relativistic language each particle of the BMV experiment is subjected to the quantum superposition of two different gravitational forces, because of the quantum split of the other particle.  In the relativistic language, different gravitational forces are due to distinct and gauge inequivalent geometries. Hence measuring the BMV effect amounts to checking that spacetime geometries can be in quantum superposition. 

We strongly hope that the BMV experiment be realised soon, not only to provide a good argument against the persistent ---in our opinion misleading--- idea of a necessarily classical spacetime, but, more excitingly, to give us a first genuine quantum gravity measurement. 

\centerline{---}

We thank Pablo Arrighi, Alessio Belenchia, Eugenio Bianchi, Sugato Bose, Chiara Marletto, Anupam Mazumdar and 
Vlatko Vedral for enlightening discussions. 

%\vfill

\bibliographystyle{JHEPs}
\bibliography{/Users/carlorovelli/Documents/library}
\end{document}